\documentstyle[prl,aps]{revtex}
\begin{document}

\title {Phonon relaxation of subgap levels in superconducting
quantum point contacts}

\author{ D. A. Ivanov$^{1,2}$ and M. V. Feigel'man$^1$}

\address{ $^1$ L.D.Landau Institute for Theoretical Physics, 117940
Moscow, Russia \\
$^2$ 12-127 M.I.T. Cambridge, MA, 02139 USA}
\date{November 2, 1998}

\maketitle

\begin{abstract}
Superconducting quantum point contacts are known to possess
two subgap states per each propagating mode. In this note we
compute the low-temperature relaxation rate of the upper
subgap state into the lower one with the emission of an acoustic
phonon. If the reflection in the contact is small, the relaxation
time may become much longer than the characteristic lifetime
of a bulk quasiparticle.
\end{abstract}

In the present paper we address the question of phonon relaxation of
subgap levels in
superconducting quantum point contacts (SQPC) 
\cite{kulik,been1,shumeiko1}.
We use this term for a
class of junctions between two superconductors (of the BCS type),
were only a small number of modes propagates and the scattering
center of the contact is much shorter than the superconducting
coherence length. Under the latter condition, the internal structure
of the scatterer is not important, but may be described by a
scattering matrix for normal electrons \cite{been2,if2}.
Thus, in the class of SQPC we
include both SNS junctions with a thin normal layer
and with quantized propagating modes \cite{takayanagi} and
mechanically-controllable breakjunctions \cite{koops}.

Theoretically, SQPCs in different setups are predicted to exhibit
various quantum phenomena. The most advanced predictions involve
quantum-mechanical evolution of a superconducting island connected
to external leads by SQPCs \cite{if1,averin4}. Other works discuss
interference between time evolutions of localized states and/or
excitations of localized states by an external electromagnetic field
\cite{gorelik1,gorelik2,shumeiko2}. For experimental observation of these
predictions, the decay time of excited localized states must be
sufficiently long. This decay time directly enters the expressions
for the thermal fluctuations of the Josephson current in SQPC
\cite{mrodero1,mrodero2,lesovik1,averin2}. Potential use of
SQPC in quantum computing devices also crucially depends on their
ability to preserve quantum coherence \cite{divinch}.

At low temperatures ($T\ll\Delta$,
where $\Delta$ is the superconducting gap), the dominant inelastic
processes are the transitions between localized subgap levels,
without participation of continuum spectrum. There
are two major channels of inelastic relaxation. One is the emission or
absorption of a phonon, the other is due to the electromagnetic
coupling to the external environment.
It will depend on the particular experimental setup, which of the
two relaxation channels dominates. In the present paper we
consider only the phonon relaxation in its simplest form, when
the superconducting
phases on the contact terminal are assumed to be rigidly fixed.

In several works, the time of decay into phonons was estimated
as the characteristic time $\tau_0$ of a bulk quasiparticle
\cite{gorelik1,gorelik2,shumeiko2}. The latter lifetime
is of order
$\tau_0 \sim \Theta_D^2 / \Delta^3$ --- the same as the relaxation
time of a normal electron with energy $\Delta$ above the
Fermi level \cite{kaplan} (we set $\hbar=1$ throughout the paper).
We want to point out that this assumption is not justified for
decay of localized states.
In refs.~\cite{averin2,bardas} the relaxation of subgap states in SQPC
is associated with bulk subgap states appearing due
to electron-phonon interaction. This contribution
is exponentially small for temperatures much lower that the
superconducting gap.

We compute the {\it direct} matrix element
for decay of localized states into phonons. This direct decay
leads to a non-vanishing relaxation rate even at zero temperature.
Further, we simplify our discussion by setting the temperature
much lower than the energy of the subgap level. Then the
only allowed process is the transition from the upper to the lower
level with the emission of an acoustic phonon. The extension of
our result to include thermal phonons is obvious (see eq.~(\ref{thermal})
below).

The characteristic energy scale of the subgap levels is $\Delta$,
thus the wavelength of the phonon is of order $ s/\Delta \sim
\xi s/v_F \ll \xi $, where $s$ and $v_F$ are sound and Fermi velocities
respectively, $\xi$ is the superconducting coherence length.
The rate and the angular distribution of the emitted phonon may depend
on the particular geometry of the contact (or, more precisely, on the
geometry of the wavefunction of the subgap states). In this paper
we discuss the simplest setup of a narrow one-dimensional contact.
By this we mean that the whole subgap state is localized in a
narrow strip of width much smaller than the phonon wavelength.
Although very idealistic, this assumption is consistent with
the model of adiabatic constriction \cite{been1} and gives
an upper bound for the actual decay rate.

Each propagating mode in a
quasi-one-dimensional contact may be described by the
Hamiltonian
\begin{eqnarray}
H=\int_{-\infty}^{+\infty} dx \Big[ &&
i\Psi^\dagger_{L\beta} \partial_x \Psi_{L\beta}
-i\Psi^\dagger_{R\beta} \partial_x \Psi_{R\beta} + \nonumber\\
&& + \Delta(x)\big(\Psi^\dagger_{R\uparrow}\Psi^\dagger_{L\downarrow}
-\Psi^\dagger_{R\downarrow}\Psi^\dagger_{L\uparrow}\big) + \nonumber\\
&& + \Delta^*(x)\big(\Psi_{R\downarrow}\Psi_{L\uparrow}
-\Psi_{R\uparrow}\Psi_{L\downarrow}\big) \Big] + H_{scatt},
\label{ham}
\end{eqnarray}
where $\Psi^\dagger$ and $\Psi$ are electron operators ($L$ and $R$
subscripts denote left- and right-movers, $\beta=\uparrow,\downarrow$ is
the spin index), $\Delta(x)$ is the superconducting gap with the following
$x$-dependence:
\begin{equation}
\Delta(x)=\cases{\Delta, & $x<0$ \cr \Delta e^{i\alpha}, & $x>0$ }
\end{equation}
[It will be convenient for us to use the units
with Fermi velocity equal to one throughout the paper].
The scattering term $H_{scatt}$ expresses elastic scattering at $x=0$ and
may be described by a scattering matrix \cite{if2}.
Diagonalizing the Hamiltonian (\ref{ham}) gives the subgap state operators
$\gamma^\dagger_\uparrow$ and $\gamma^\dagger_\downarrow$ raising
energy by
\begin{equation}
E(\alpha)=\pm \Delta\sqrt{1-t\sin^2{\alpha\over 2}}
\label{levels}
\end{equation}
each ($t$ is the normal transparency of the contact)\cite{been2,furusaki}.
The two levels below continuum are the ground state $|0\rangle$
and the first excited state $|1\rangle = \gamma^\dagger_\uparrow
\gamma^\dagger_\downarrow |0 \rangle$. The decay of the state $|1\rangle$
to the state $|0\rangle$ with the emission of a phonon depends on the
density matrix element
\begin{equation}
\langle 0| n(x) |1\rangle =
\langle 0| \Psi^\dagger_\beta(x)
\Psi_\beta(x) | 1\rangle.
\label{density}
\end{equation}
Since
$\gamma^\dagger_\uparrow$ and $\gamma^\dagger_\downarrow$
are linear in electron operators, the matrix element
(\ref{density}) may be computed by commuting density operator
with them:
\begin{equation}
\langle 0| n(x) |1\rangle =
\left\{ \left[ n(x),\gamma^\dagger_\uparrow \right],
\gamma^\dagger_\downarrow \right\}= i |b|\kappa e^{-2\kappa |x|}
{\rm sign}(x),
\label{density2}
\end{equation}
where $b$ is the backscattering amplitude ($|b|=\sqrt{1-t}$),
and $\kappa=\sqrt{t}\Delta|\sin(\alpha/2)|$ is the inverse
length of the subgap state. The matrix element is purely imaginary
if the relative phases of $|0\rangle$ and $|1\rangle$ are chosen
according to \cite{if2}
\begin{equation}
\langle 0 | {\partial \over
 \partial\alpha} |0\rangle =
\langle 1 | {\partial \over \partial\alpha} |1\rangle = 0,
\qquad
\langle 0 | {\partial\over\partial\alpha} |1\rangle {\rm ~is~real}.
\label{real}
\end{equation}
This fact is of no importance for the present calculation, but will
be used elsewhere in the discussion of the phonon emission in the
presence of dynamics in $\alpha$.

The electron-phonon interaction is described by the deformation potential:
\begin{equation}
H_{e-ph} = g\int d^3 r\, \varphi(r) \Psi^\dagger_\beta(r)
\Psi_\beta(r),
\end{equation}
\begin{equation}
\varphi(r)={1\over\sqrt V} \sum_k \sqrt{\omega_k\over 2}
(b_k e^{i(kr-\omega_k t)} + b_k^\dagger e^{-i(kr-\omega_k t)}),
\end{equation}
where $b_k$ are phonon operators normalized by $[b_{k_1},
b^\dagger_{k_2}] = \delta_{k_1 k_2}$,
\begin{equation}
g^2={\pi^2 \zeta \over 2 \varepsilon_F},
\end{equation}
(in the units with Fermi velocity equal to one), $\zeta$ is the coupling
constant of order one.

The transition rate is then given by
\begin{equation}
\tau^{-1}=2\pi\sum_k \Big| \langle 0,k | H_{e-ph} | 1\rangle \Big|^2
\delta(\omega_k-2E)=
\pi^3\zeta {E\over \varepsilon_F^2} \int {d^3k\over (2\pi)^3}
\delta(\omega_k-2E) \Big| \langle 0 | n_k | 1 \rangle \Big|^2.
\end{equation}
Here $E$ is the energy of the subgap states given by (\ref{levels})
[so that the energy of the emitted phonon is $2E$], $\varepsilon_F$
is the Fermi energy, $\omega_k$ is the phonon dispersion relation,
and $n_k$ is the three-dimensional density operator.
Assume the linear isotropic phonon spectrum:
\begin{equation}
\omega_k=s|k|
\end{equation}
and the narrow contact limit, where the matrix element of $n_k$
depends only on the component of $k$ parallel to the constriction
and is given by the Fourier transform of the one-dimensional
matrix element (\ref{density2}). Finally, using $k\gg\kappa$,
we arrive to the answer for $\tau^{-1}$:
\begin{equation}
\tau^{-1}=\pi^2\zeta {E^2\over (c\varepsilon_F)^2} \int_{-\infty}^\infty
dx\, \Big|\langle 0 | n(x) |1 \rangle \Big|^2
={\pi^2\zeta\over 2} (1-t) {E^2\kappa\over (s\varepsilon_F)^2}.
\end{equation}
Returning to the physical units, we find up to a constant factor
of order one
\begin{equation}
\tau^{-1}=\sqrt{t}(1-t)\Big|\sin{\alpha\over 2} \Big|
(1-t \sin^2{\alpha\over 2}) {\Delta^3\over \Theta_D^2},
\label{result}
\end{equation}
where $\Theta_D$ is the Debye temperature.

If compared to the characteristic bulk quasiparticle inverse lifetime
$\tau_0^{-1} \sim \Delta^3/\Theta_D^2$ \cite{kaplan},
the result (\ref{result}) is
smaller by a factor depending on the backscattering probability
 $1-t$\cite{comment}.
In the case of weak backscattering, this factor may contribute up to
orders of magnitude to the decay time. This effect is easy to
understand: in ideally conducting contact ($t=1$) the two Andreev states
carry opposite momenta equal to the Fermi momentum, and the matrix
element (\ref{density}) contains only a rapidly oscillating with
momentum $2k_F$ part \cite{umklapp}:
\begin{equation}
\langle 0 | n(x) |1\rangle = \kappa e^{-2\kappa |x|}
e^{2ik_F x}.
\end{equation}
This oscillating part of the matrix element gives the lower bound
for the relaxation rate (\ref{result}) as $t \to 1$:
\begin{equation}
\tau^{-1}_{t=1}
\sim {E^3 \kappa^4 \over s^3 \varepsilon_F^6}
\sim \left|\cos^3\left({\alpha\over 2}\right)\right| 
\sin^4\left({\alpha\over2}\right)
{\Delta^7\over \varepsilon_F^3 \Theta_D^3}.
\label{result2} 
\end{equation}
For realistic values of $\Delta$, $\Theta_D$, and $\varepsilon_F$,
this relaxation time is unphysically large. The actual relaxation time
at $t=1$ will be bounded by other factors such as finite thickness of the
interface and non-one-dimensionality of the contact. These effects go
beyond the simple model of the present paper \cite{lg}.

Another important feature of
the relaxation rate (\ref{result}) is that it vanishes
at $\alpha\to 0$, since in this limit the subgap states become
delocalized.

Our assumption of one-dimensionality of the contact also results
in overestimating the relaxation rate. If the ``tails'' of the
localized states are smeared in the terminals, they give weaker
contribution to phonon emission. Unfortunately, this effect is highly
geometry-dependent, and should be considered separately in each
experimental realization.

As a direct application of the above result, the
decay rate entering the fluctuations of the Josephson
current in a single SQPC \cite{mrodero1,mrodero2,averin2}
is given by
\begin{equation}
\gamma=\tau^{-1} (1+2 n_B(2E(\alpha))=\tau^{-1} \coth{E(\alpha)\over T},
\label{thermal}
\end{equation}
where $n_B(2E(\alpha))$ is the Bose occupation number for the phonons
involved in the transition \cite{lesovik1}, $\tau^{-1}$ is the
zero-temperature rate (\ref{result}). The low-frequency current
noise is \cite{mrodero1,averin2}
\begin{equation}
S(\omega)=S_0 {2\gamma\over \omega^2+\gamma^2},
\end{equation}
where $S_0$ is the integral low-frequency noise.

To summarize, we calculated the rate of direct relaxation of subgap
states in SQPC into acoustic phonons at low temperature in the
simplest one-dimensional geometry. The relaxation rate does not vanish
in the $T \to 0$ limit, but it is strongly suppressed in the
case of a nearly ballistic contact.

We thank Gordey Lesovik for helpful discussions.
The research of M.V.F. was supported by INTAS-RFBR grant \#
95-0302, Swiss National Science Foundation collaboration grant \# 7SUP
J048531, DGA grant \# 94-1189, and the Program "Statistical Physics"
from the Russian Ministry of Science.


\begin{references}
\bibitem{lg}
Our results (\ref{result}), (\ref{result2})
disagree with ref.~\cite{lesovik1} where the
relaxation rate is computed in the same approximation. The relaxation rate
of ref. \cite{lesovik1} is not suppressed
in the ballistic limit $t \to 1$.
\bibitem{umklapp}
We neglected the possibility of umklapp scattering.
\end{references}
\end{document}